\documentclass[
reprint,
floatfix,
aps,
prb,
amsmath,
twocolumn,
superscriptaddress,
amssymb,
tightenlines,
groupaddress,
footinbib
]{revtex4-2}

\usepackage{amsfonts,amssymb}
\usepackage[subnum]{cases}
\usepackage{mathrsfs}
\usepackage{amsmath}
\usepackage[none]{hyphenat}
\usepackage{hyperref} 
\usepackage{bm} 
\usepackage{natbib}
\usepackage{soul} 
\usepackage{color}
\usepackage{longtable}
\usepackage{ textcomp }
\usepackage{verbatim}
\usepackage{fancyhdr} 
\usepackage{lastpage} 
\usepackage{extramarks} 
\usepackage{graphicx} 
\usepackage{lipsum} 
\usepackage{amssymb}
\usepackage{slashed}
\usepackage{tikz}
\usepackage{comment}
\usepackage{natbib}
\usepackage[caption=false,listofformat=parens, subrefformat=parens]{subfig}
\usepackage{marvosym}
\usepackage{array}

\usepackage{amsthm} 

\newcommand{\abs}[1]{\left| #1 \right|} 
 
\let\baraccent=\= 
\renewcommand{\=}[1]{\stackrel{#1}{=}} 

\theoremstyle{definition}

\theoremstyle{remark}

\newcolumntype{C}[1]{>{\centering\let\newline\\\arraybackslash\hspace{0pt}}m{#1}}

\usepackage{physics}
\usepackage{CJKutf8}

\newcommand{\lrp}[1]{\left( #1 \right)}
\newcommand{\lrb}[1]{\left[ #1 \right]}

\begin{document}

	\title{Broadband Focusing of Acoustic Plasmons in Graphene with an Applied Current}
	
	\author{Michael Sammon}
	\email{sammo017@umn.edu}
	\affiliation{Department of Electrical and Computer Engineering, University of Minnesota, Minneapolis, MN 55455, USA}
	\author{Dionisios Margetis}
	\affiliation{Institute for Physical Science and Technology, and Department of Mathematics, and
Center for Scientific Computation and Mathematical Modeling, University of Maryland, College Park, MD 20742, USA}
	\author{E. J. Mele}
	\affiliation{Dept. of Physics and Astronomy, University of Pennsylvania, Philadelphia, Pennsylvania 19104, USA}
	\author{Tony Low}
	\email{tlow@umn.edu}
	\affiliation{Department of Electrical and Computer Engineering, University of Minnesota, Minneapolis, MN 55455, USA}
	\date{\today}
	
	\begin{abstract}
Non-reciprocal plasmons in current-driven, isotropic, and homogenous graphene with proximal metallic gates is theoretically explored. Nearby metallic gates screen the Coulomb interactions, leading to linearly dispersive acoustic plasmons residing close to its particle-hole continuum counterpart. We show that the applied bias leads to spectral broadband focused plasmons whose resonance linewidth is dependent on the angular direction relative to the current flow due to Landau damping. We predict that forward focused non-reciprocal plasmons are possible with accessible experimental parameters and setup. 
\end{abstract}

	\maketitle

	\begin{figure}[h]
		\includegraphics[width=\linewidth]{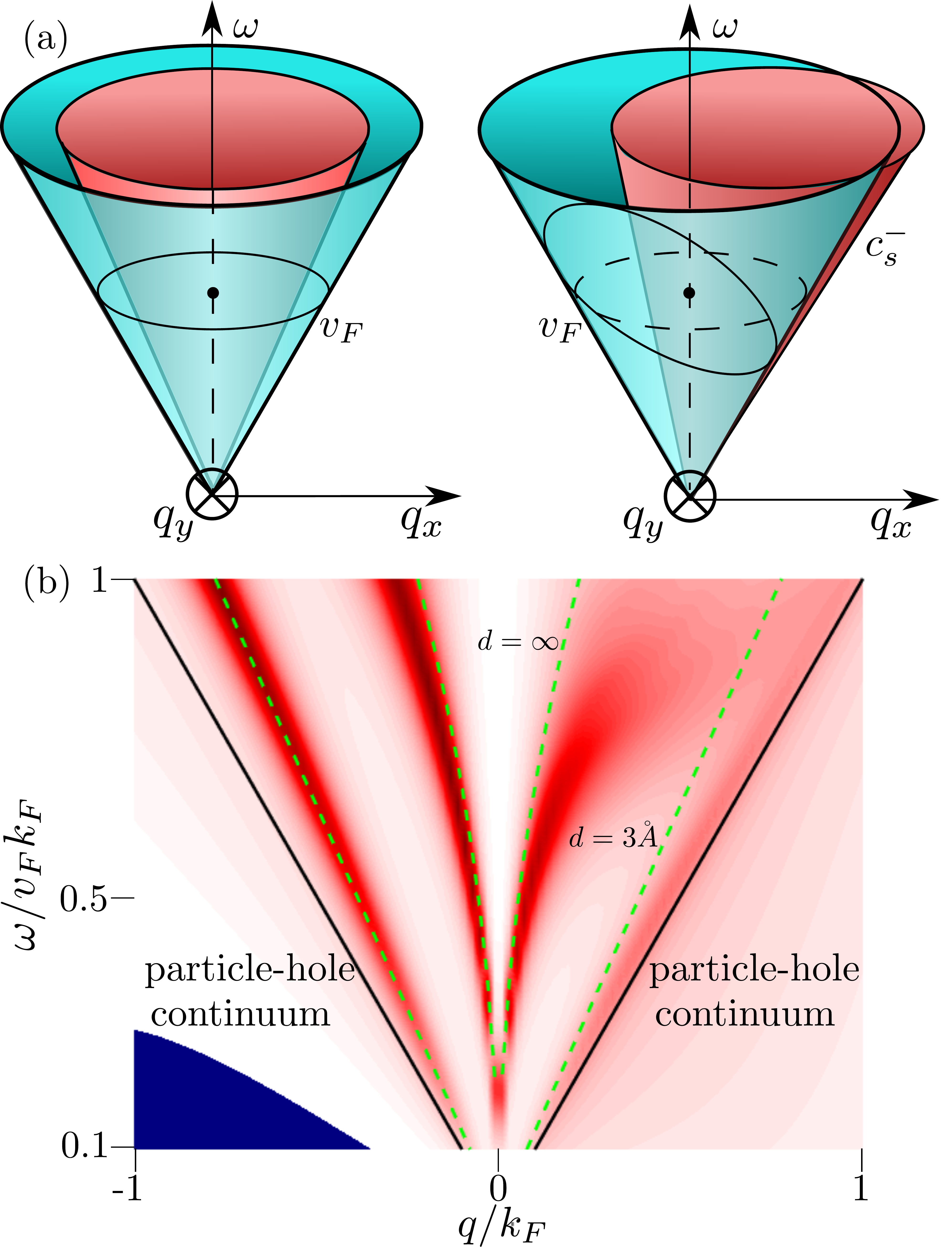}\label{fig:cartoon}
		\caption{(a)Cartoon illustration of the acoustic plasmon dispersion in graphene. 
		Left: Acoustic Plasmon dispersion in graphene with no current bias. Right:Acoustic Plasmon dispersion in graphene with a current bias in the positive $x$ direction. The plasmon dispersion is illustrated by the red cone, while the teal cone is the onset of the particle-hole continuum. The acoustic plasmon cone is tilted in the direction of the applied bias, resulting in the shift of the plasmon dispersion. Solid circles illustrate the Fermi energy. For comparison, we have included the Fermi energy at zero bias as a dashed line in the right plot. 
		(b)Colormap of the loss function $S(\vb{q},\omega)=\lrb{1-U(\vb{q})\Pi(\vb{q},\omega)}^{-1}$ for a graphene-insulator-metal system with an applied current obtained by numerical integration. The blue shows a region where gain is achievable. Both the current and the wave number $q$ are assumed to be in the $x$-direction. Calculations were made using $E_F=0.2$\,eV, $k_s=0.6k_F$, and $\kappa=1$. Both the acoustic plasmon for $d=3$\AA\, and the plasmon for $d\rightarrow\infty$ are shown. For comparison we have included the analytical dispersion at zero bias shown by the green lines. }
	\end{figure}

As optoelectronic technology advances, there is increasing demand for the development of strongly non-reciprocal light based devices. \cite{koenderink_2015,lira_2012,fan_2012,sounas_2013} While magneto-optical systems are well known for exhibiting non-reciprocity, the need for an external magnetic bias makes it difficult to integrate these systems onto nanophotonic platforms. This has driven a shift of the focus in recent years to the use of an external current bias to induce non-reciprocity.\cite{Morgado_Uni,Morgado_2020,Bliokh_2018,Stauber_Bias,Levitov_Doppler,Levitov_Flow,Dyakonov_Shur,Van_2016} Particular focus has been given to graphene surface plasmons, as their wide frequency range and gate tunability offer a wide variety of applications.\cite{DasSarma,Stauber_Graphene,Acoustic,Low_2014,Wunsch_2006,Fei_2012,Chen_2012}

The effect of a current bias on graphene surface plasmons is to induce a Doppler shift in the plasmonic spectrum. By this effect, plasmons that are moving in the same direction as the electron flow are blue-shifted, while plasmons moving against the electron flow are red-shifted.\cite{Levitov_Doppler} It has been shown that when the drift velocity is a significant fraction of the Fermi velocity in graphene, there exists a band of frequencies in which the red-shifted plasmons are forbidden and the plasmon is effectively unidirectional.\cite{Levitov_Doppler,Morgado_Uni,Morgado_2020} 

In this Letter, we show that a more versatile phenomenon can be found in graphene-dielectric-metal systems which exhibit acoustic plasmons.\cite{Lee_2019,Alonso_2017,Lundeberg_2017,Menabde_2021} When the separation $d$ between the metal and graphene is small, the sound velocity of the acoustic plasmon is within a few percent of the Fermi velocity; and the plasmon branch lies in close proximity to the particle-hole continuum.\cite{Acoustic,Stauber_2012,DasSarma_2009,Voronin_2020} If in addition an external bias is applied to the system, the plasmons moving against the electron flow can be brought into the particle-hole continuum and become damped. This focuses the acoustic plasmons in the direction of the electron flow. Below we provide analytical and numerical calculations which show how the focusing effect depends on the applied current, the electron density, and the dielectric environment. We emphasize that the major advantage in this system compared to the conventional plasmons is that the focusing effect occurs at all frequencies. Hence, it is more suitable for broadband applications.

	Within the random phase approximation (RPA), the dynamical polarization of graphene is given by\cite{Stauber_Graphene,DasSarma}
	\begin{align}\label{eq:Pol_Full}
	\Pi(\vb{q},\omega)=\frac{g_sg_v}{(2\pi)^2}\sum_{s,s'}\int d^2\vb{k} f_{s,s'}(\vb{k},\vb{q})\\
	\times\frac{n_F[E^{s'}(\vb{k}+\vb{q})]-n_F[E^{s}(\vb{k})]}{E^{s'}(\vb{k}+\vb{q})-E^{s}(\vb{k})-\hbar\omega-i0^{+}},
	\end{align}
	where $g_{s(v)}=2$ is the spin(valley) degeneracy, the prefactor $f_{s,s'}(\vb{k},\vb{q})$ is the band-overlap integral, and $E^s(\vb{k})$ is the dispersion of the conduction $(s=1)$ and valence band $(s=-1)$. We assume that the graphene is isotropic and spatially homogenous. Within the Dirac-cone approximation, we have $E^s(\vb{k})=s\hbar v_Fk$ with $v_F\simeq 9\times10^7$ cm\,s$^{-1}$, and 
	\begin{equation}
	f_{s,s'}(\vb{k},\vb{q}) = \frac{1}{2}\lrp{1+ss'\frac{k+q\cos(\theta_k-\theta_q)}{\abs{\vb{k}+\vb{q}}}},
	\end{equation} 
	where $\theta_p$ is the angle between the wave vector ${\bf p}$ (${\bf p}= {\bf k}$ or ${\bf q}$) 
	and the $x$-axis. The equilibrium electronic occupation is determined by the Fermi-Dirac distribution 
	\begin{equation}
	n_F(E)=\lrb{\exp\lrp{\frac{E-E_F}{k_BT}}+1}^{-1}
	\end{equation}
	where $E_F=\hbar v_Fk_F$ is the Fermi energy measured relative to the charge neutrality point, and $T$ is the temperature of the system. We assume for definiteness that $E_F>0$.

	In the presence of an applied current, the electrons reach a new quasi-equilibrium with a modified distribution function $n_F^*(E)$. The modified Dirac distribution can be modelled by shifting the Fermi surface of the biased electrons by an amount $\vb{k_s}=-e\tau\vb{j}/(\hbar\sigma)$, where $\vb{j}$ is the current density, $\sigma$ is the conductivity in graphene, and $\tau$ is the transport scattering time.\cite{Stauber_Bias,Levitov_Doppler} Assuming that the current is in the positive $x$-direction, the Fermi wave number obtains an angular dependence described by
	\begin{equation}
	k_F^*(\theta) = -k_s\cos\theta+\sqrt{k_F^2-k_s^2\sin^2\theta},
	\end{equation} 
	where we assume $k_s<k_F$ and $\theta=\theta_k-\theta_q$. An illustration of the shifted Fermi surface is shown in Fig.\,1(a). Assuming that the Fermi surface is shifted in the positive $x$-direction, we see that the shifted Fermi surface depletes the electron concentration in the direction of the current and increases their concentration in the opposite direction. As we argue below, this shift of the Drude weight is responsible for the red-shift of the dispersion in the direction of the current. 
	
	The plasmon dispersion $\omega(q)$ is determined by solving the equation
	\begin{equation}\label{eq:Plasmon}
	1=U(\vb{q})\Pi(\vb{q},\omega),
	\end{equation}
	where 
	\begin{equation}\label{eq:Coulomb}
	U(\vb{q})=\frac{e^2}{2\kappa \varepsilon_0 q}\lrb{1-\exp(-2qd)}\approx\frac{e^2d}{\kappa\varepsilon_0} 
	\end{equation}
	is the Coulomb interaction of the graphene-dielectric-metal system, $\kappa$ is the dielectric constant of the system, and $d$ is the thickness of the dielectric layer, and we have used the approximation $qd\ll1$. In order to obtain analytical results, we consider the limit $v_Fk_s\ll v_Fq$, and  $\omega\ll v_Fk_F$. In this limit, the band overlap integral simplifies to $f_{s,s'}(k,q)=\frac{1}{2}(1+ss')$, allowing us to focus only on the intraband contribution. As we show in the Supplemental material, for $q$ along the $x$-direction $\Pi(q,\omega)$ takes a surprisingly simple form,
	\begin{gather}\label{eq:Pol_exp}
	\Pi(\vb{q},\omega)=D(E_F)\lrb{\Pi_0+\delta\Pi_1+\delta\Pi_2},\\
	\Pi_0(\vb{q},\omega) = -\lrb{1-\frac{c_s}{\sqrt{c_s^2-1}}},\label{eq:Pol0}\\
	\delta\Pi_1(\vb{q},\omega)=\frac{k_s}{k_F}\cos\theta_q\lrb{2c_s+\frac{1-2c_s^2}{\sqrt{a^2-1}}},\label{eq:Pol1}\\
	\delta \Pi_2(\vb{q},\omega) =\frac{1}{2}\lrp{\frac{k_s}{k_F}}^2\biggr[\cos^2\theta_q\left(\frac{3}{2}-3c_s^2+3c_s\sqrt{c_s^2-1}\right)-...\notag \\ 
	...-\sin^2\theta_q\left(\frac{1}{2}-3c_s^2+\frac{3c_s^3-c_s}{\sqrt{c_s^2-1}}\right)\biggr],\label{eq:Pol2}
	\end{gather}
	where $c_s=\omega/(v_Fq)$, and $D(E_F)=g_sg_vE_F/[2\pi(\hbar v_F)^2]$ is the density of states in graphene. In Eq.\,(\ref{eq:Pol_exp}) we have included terms up to second order in $k_s/k_F$. Thus we find the acoustic plasmon dispersion by substituting Eqs.\,(\ref{eq:Pol_exp})-(\ref{eq:Pol2}) into Eq.\,(\ref{eq:Plasmon}). Expanding $c_s=c_{s0}+\delta c_{s1}+\delta c_{s2}$ we find
	\begin{gather}
	c_{s0}=\frac{1+A}{\sqrt{1+2A}},\label{eq:cs0}\\
	\delta c_{s1} = -\frac{k_s}{k_F}\cos\theta_q(c_{s0}^2-1)^{3/2}\lrb{\frac{2c_{s0}^2-1}{\sqrt{c_{s0}^2-1}}-2c_{s0}},\label{eq:cs1}\\
	\delta c_{s2}=-\frac{1}{2}\lrp{\frac{k_s}{k_F}}^2(c_{s0}^2-1)^{3/2}\biggr[\cos^2\theta_q\lrp{3c_{s0}^2-\frac{3}{2}-3c_{s0}\sqrt{c_{s0}^2-1}}-...  \notag  \\
	-\sin^2\theta_q\left(\frac{1}{2}-3c_{s0}^2+\frac{3c_{s0}^3-c_{s0}}{\sqrt{c_{s0}^2-1}}\right)\biggr],\label{eq:cs2}
	\end{gather}
	 where $A = (e^2dD(E_F))/\kappa/\varepsilon_0$, and we have used Eq.\,(\ref{eq:Coulomb}) in the limit of small $q$. 
	 \begin{figure}
	 	\includegraphics[width=\linewidth]{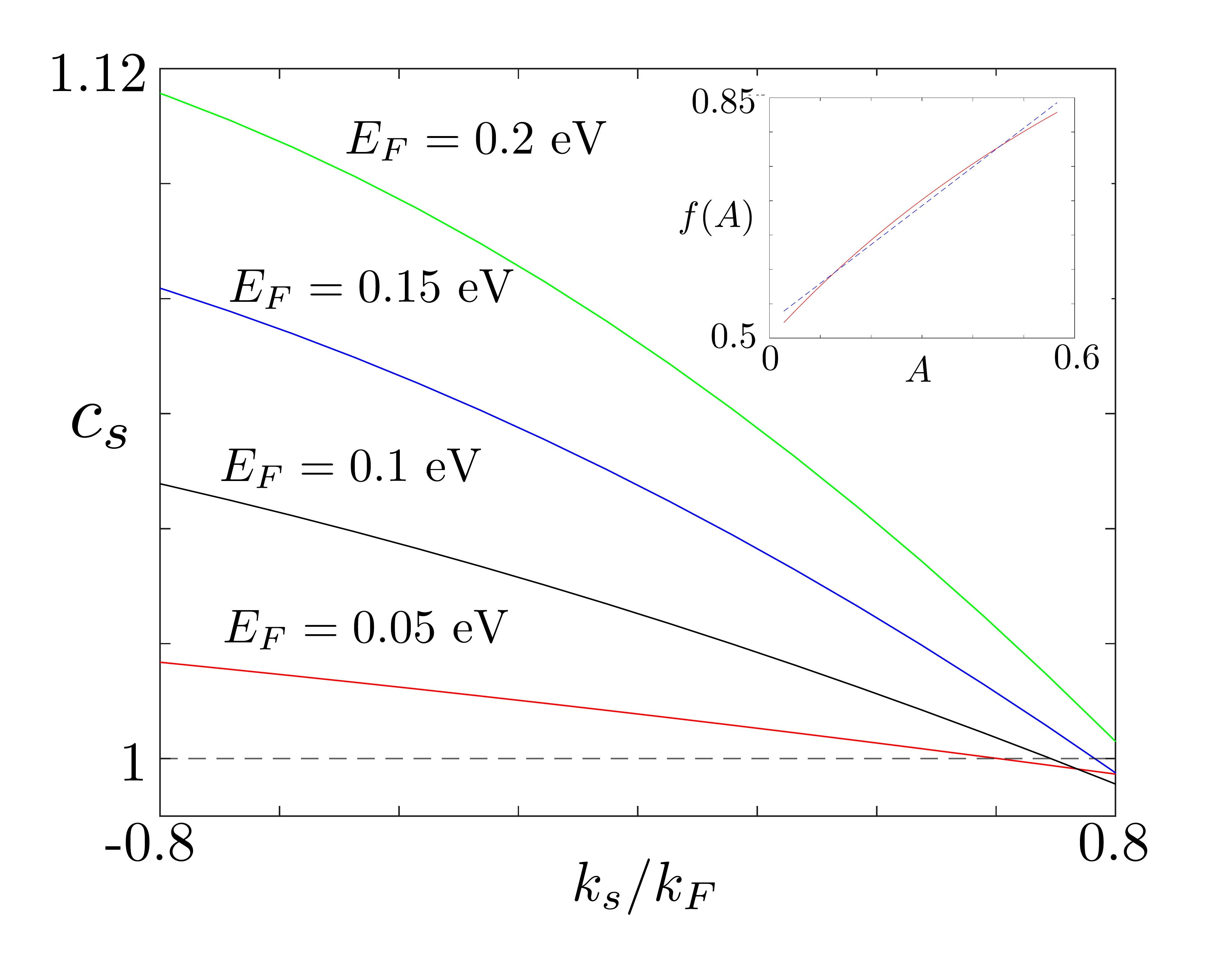}
	 	\caption{Dimensionless acoustic plasmon velocity $c_s$ as a function of the shifted Fermi surface center $k_s$ for several different values of $E_F$. Negative values of $k_s$ correspond to the $q<0$ branch. Inset:The dimensionless function $f(A)\equiv k_{sc}/k_F$ vs the dimensionless parameter $A$. The range of $A$ is for $0.01\leq E_F \leq 0.20$ and for the dielectric constant $\kappa =3$. The dashed line in the inset is the best linear fit. }\label{fig:cs}
	 \end{figure}
	 
	 Let us discuss the dispersion defined by Eqs.\,(\ref{eq:cs0})-(\ref{eq:cs2}), which is plotted in Fig.\,\ref{fig:cs}.  As we see from Eq.\,(\ref{eq:cs1}), the dependence on $\cos\theta_q$ creates a non-reciprocity between the $q<0$ and $q>0$ branches.  As the current increases, the $q>0$ $(q<0)$ plasmon branch is red-shifted (blue-shifted) relative to the zero current value defined in Eq.\,(\ref{eq:cs0}). This effect creates a particularly interesting situation for acoustic plasmons in graphene, as it suggests that there is some $k_{sc}$  above which $c_s<1$.  Under this condition, the $q>0$ plasmon branch  \emph{lies entirely in the Landau damping region} and is strongly damped, while the $q<0$ branch is very weakly attenuated (see Fig.\,1\,(a)). In this regime the plasmon is focused in the opposite direction of the applied bias. We can solve for $k_{sc}$ using Eqs.\,(\ref{eq:cs0})-(\ref{eq:cs2}) and define
	 \begin{equation}\label{eq:k_sc}
	     \frac{k_{sc}}{k_F}=f(A),
	 \end{equation}
	 where $f(A)$ is a dimensionless function that monotonically increases with the parameter $A$. This function is shown as the inset of Fig.\,(\ref{fig:cs}) for $\kappa =3$ and $E_F$ ranging from 0.01-0.2 eV. 
	 
	 Eq.\,(\ref{eq:k_sc}) has some important experimental implications. The wave number $k_{sc}$ defines a critical current
	 \begin{equation}\label{eq:j_sc}
	     j_{sc}=ev_Fnf(A),
	 \end{equation}
	in which the acoustic plasmon focusing occurs. Here $n\propto E_F^2$ is the electron density. By consideration of $A\propto E_F/\kappa$,
	Eq.\,(\ref{eq:j_sc}) suggests that $j_{sc}$ can be tuned by either changing the density with an external gate, or changing the dielectric environment of the sample. As an example, let us consider what this means for hBN encapsulated graphene which has some of the highest saturation currents measured in graphene. For an electron density of $7\times10^{11}$ cm$^{-2}$ and the dielectric constant $\kappa=3.29$,\cite{Dielectric_hBN} the critical current density is approximately 800 A/m. This suggests that the focusing of the acoustic plasmon can be observed at these densities.\cite{Velocity_Saturation}

	\begin{figure}
		\includegraphics[width=\linewidth]{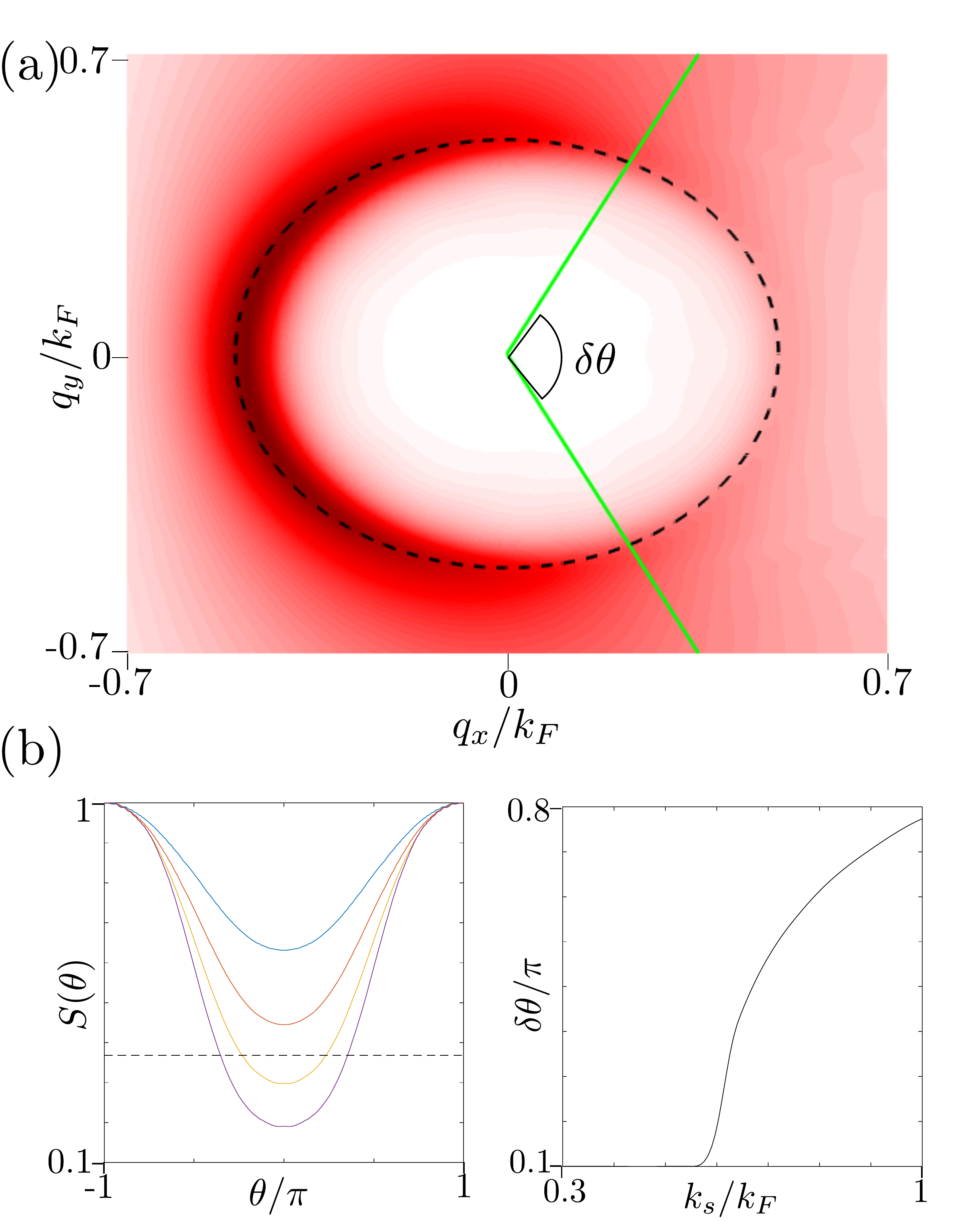}
		\caption{(a)Temperature plot of the loss function $S(\vb{q},\theta)$ for $E_F=100$meV and $\kappa=3$ at a fixed frequency $\hbar\omega=0.5E_F$ for $k_s=0.9k_F$. The dashed black line illustrates the boundary of the particle-hole continuum. The green line shows the angle $\delta\theta$ in which $S(q,\theta)$ is depleted.
		(b)Left: Plot of the isofrequency contour $S(\theta)$ for $E_F = 100$\,meV, $\kappa=3$, and $\hbar\omega=0.5E_F$ at several $k_s=0.3k_F$ (blue), $k_s=0.5k_F$ (red), $k_s=0.7k_F$ (orange), and $k_s=0.9k_F$. Each line is normalized to the maximum value of $S(\theta)$ at fixed $k_s$. The dashed line shows the value we use to define the depletion angle $\delta\theta$.
	 	Right: Plot of the range $\delta\theta$ of angles around the current bias direction in which the isofrequency contour $S(\theta)$ is exponentially depleted as function of the relative shift of the Fermi surface center. Plasmons emitted at this bias are focused in a range $2\pi-\delta\theta$ opposite the applied current bias. }
	\end{figure}

	 Let us now compare these results to those obtained by a numerical calculation of the polarization defined by Eq.\,(\ref{eq:Pol_Full}). In Fig.\,3(a) we present a slice of the loss function $S(q,\theta)$ at a fixed frequency $\hbar\omega=0.5E_F$ for $E_F=100$meV and $\kappa=3$  for $k_s=0.9k_F$. Here we use the polar coordinates $q$ and $\theta$ for describing features of $S(q,\theta)$, and drop $\omega$ from the argument for simplicity of notation. The particle-hole continuum is illustrated by the dashed black circle. There are several essential features to notice. First, let us define the isofrequency contour $S(\theta)$ as the maximum of $S(q,\theta)$ at a fixed $\theta$. Unlike the zero bias case, it is clear that the isofrequency contour has shifted due to the bias, and is no longer circularly symmetric. More importantly, a range of angles $\delta\theta$ around $\theta =0$ has crossed into the particle-hole continuum. The spectral intensity of this region has been nearly fully depleted. In order to investigate this more thoroughly, we plot the isofrequency contour $S(\theta)$ in the left plot of Fig.\,3(b) . The values are normalized to the maximum value of $S(\theta)$ for each value of $k_s$ separately. We see that as the current bias is increased, the minimum of $S(\theta)$ decreases and the size of the depleted region grows, suggesting that as the bias is increased the acoustic plasmon becomes more focused in the direction opposing the current. This allows us to define a measure of the focusing by the range of angles $\delta\theta$ in which the loss function $S(\theta)<\text{max}(S(\theta))/\gamma$, where $\gamma$ is Euler's constant and is presented in the right plot of Fig.\,3(b). By this criterion, the acoustic plasmon focusing does not occur until a minimum value of $k_s\approx 0.6k_F$ is applied at this Fermi energy. 
	 
	 \begin{figure}
	     \centering
	     \includegraphics[width = \linewidth]{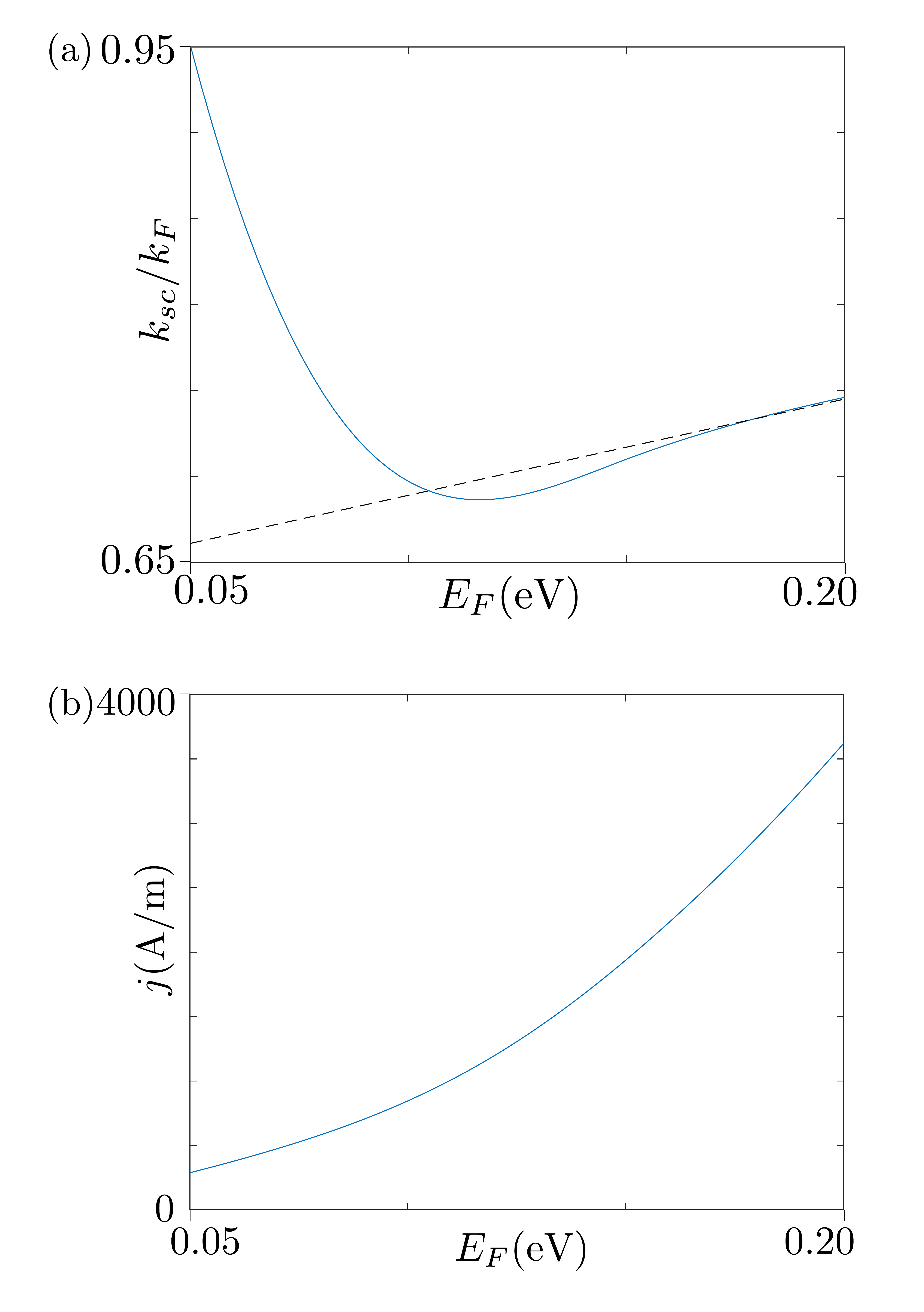}
	     \caption{(a) The critical value $k_{sc}$ of the shift of the Fermi surface center vs. the Fermi energy $E_F(eV)$. $k_{sc}$ is defined as the minimum shift in which the angle of depletion $\delta\theta>\pi/2$. The dashed line is the best linear fit of the large $E_F$ dependence, given in the text.
	     (b) The critical current $j_{sc}$ vs. the Fermi energy $E_F$. The values of $j_{sc}$ are obtained from $k_{sc}$ according to Eq.\,(\ref{eq:j_sc}). }
	     \label{fig:j_sc}
	 \end{figure}
	 
	 Let us discuss how these results can be tuned with the (unbiased) Fermi energy. From the angular dependence of the loss function, we can define a critical value $k_{sc}$ above  which the angle of depletion $\delta\theta>\pi/2$. In Fig.\,\ref{fig:j_sc} we plot $k_{sc}$ vs $E_F$ based on this criterion, along with the associated current obtained from Eq.\,(\ref{eq:j_sc}). At larger $E_F$, there is an approximately linear dependence of $k_{sc}$ on the Fermi energy $k_{sc}/k_F=ME_F+B$, with $M=0.56$(eV)$^{-1}$ and $B=0.6$, as shown by the dashed line in Fig\,\ref{fig:j_sc}(a). This should be compared with our original analytical calculation in which $M= 1.6$(eV)$^{-1}$ and $B=0.6$. We note that our analytical expression dramatically overestimates the slope. This is likely due to the difference in definition of $k_{sc}$ between the analytical and numerical calculations. The analytical calculation is determined by the plasmon dispersion along the $x$-axis, which is most sensitive to the applied bias and thus more sensitive to changes in the Fermi energy and system parameters.  At smaller Fermi energies, there is a sudden increase in the value of $k_{sc}$. This is likely due to the large intrinsic broadening of the plasmon peak of $S(q,\omega)$ at smaller Fermi energies. As the Fermi energy is decreased, the plasmon peak at zero bias is pushed closer to the particle hole continuum, eventually having a partial overlap due to the finite width of the plasmon peak. As the focusing effect is caused by the red-shift of the plasmon dispersion into the particle-hole continuum with a finite bias, this partial overlap at zero bias weakens the effect.\cite{Ni_2018}

	 We briefly comment on the effects of disorder on these results. For hard scattering centers, the large momentum transfer makes it possible to couple opposing plasmon modes. It is reasonable to worry that this coupling would cause significant damping of the blue-shifted branch, destroying the focusing effect. However, it has been shown in previous studies that the main impact of disorder scattering on the plasmon damping factor is through the coupling of the plasmon to the particle hole continuum.\cite{Polini_Intrinsic,Polini_Disorder} For moderate disorder this increases the damping by roughly a factor of 2 and weakens the proposed effect. However, with the unprecedented advancement in high quality graphene heterostructure fabrication, experimental realization of our predictions in high quality factor plasmon in graphene should be within reach. 
	 
	 Before concluding, let us briefly dwell on the region of Fig.\,1(b) in which there is gain, i.e. the region of negative loss. As we show in Fig.\,1(a), the applied bias causes a population inversion in which there is a concentration of higher energy electrons in the direction opposing the current, and lower energy empty states in the direction of the current. The driving of the particles from the high energy occupied states into the lower energy states results in an energy gain in the system, and has been studied previously in optically pumped systems.\cite{Winzer_2013} In the systems being studied here, this effect is limited to large wave numbers and low frequencies, limiting its potential use.\cite{Ni_2018}  
	 
	 To summarize, we have studied the effects of an applied current bias on acoustic plasmons in metal-dielectric-graphene systems. We have shown that for thin dielectric layers, the applied current bias focuses the acoustic plasmons in the direction opposite to the current bias. This focusing effect can be enhanced by depleting the electron concentration, or by using a material with a higher dielectric constant. We emphasize that due the linear dispersion of the acoustic plasmons, the focusing effect is spectrally broad, making it ideal for development of nonreciprocal light based devices. 
	 
	 \emph{Acknowledgment} M.S, E.M. and T.L. acknowledge support from the National Science Foundation under Grant No. NSF/EFRI-1741660. D.M. acknowledges partial support by the ARO MURI Award No. W911NF-14-1-0247.

 \bibliography{Plasmon}

\end{document}


\section{Biased Acoustic Plasmon Dispersion}
Here we would like to find an analytical estimate of the graphene acoustic plasmon dispersion in the presence of an applied current bias. We begin with the polarization defined by Eq.\,(1) in the main text. We assume that the wave number $\vb{q}$ and frequency $\omega$ satisfy $v_F q\sim\omega\ll v_Fk_F$. Under this assumption, the polarization is entirely determined by the intraband contribution ($s=s'=1$), and the band overlap integral $f_{s,s'}(\vb{k},\vb{q})=1$.  After a few simplifications, it is easy to show that $\Pi(\vb{q},\omega)$ can be rewritten as 
\begin{gather}
	\Pi(\vb{q},\omega)=D(E_F)\int_0^{2\pi} \frac{d\theta_k}{2\pi} I(\theta_k),\label{eq:Pol_simple}\\
		I(\theta)=\sum_{b=\pm 1}\int_0^{Q_c(\theta)}dx\frac{x}{x-\sqrt{x^2+y^2+2byx\cos(\theta)}+bz},\label{eq:I_theta}
\end{gather}
	where as in the main text we use $\theta = \theta_k-\theta_q$, $D(E_F) = g_sg_vE_F/[]2\pi(\hbar v_F)^2]$, is the density of states and we have used the dimensionless variables $x = k/k_F, y = q/k_F$, and $z = \omega/v_F k_F$. We have also introduced the dimensionless shifted Fermi wave number 
	\begin{equation}
		Q_c(\theta_k) = -x_s\cos\theta_k+\sqrt{1-x_s^2\sin^2\theta_k},		
	\end{equation}
	where $x_s=k_s/k_F$.
		 The integral in Eq.\,(\ref{eq:I_theta}) can be determined through two substitutions. First we make the substitution 
	\begin{equation}\label{eq:s_def}
		x +by\cos(\theta) = y\abs{\sin(\theta)}\sinh(s),
	\end{equation} 
which transforms the integral to 
\begin{gather}\label{eq:s_integral}
	\frac{q^2\abs{\sin(\theta)}}{2}\int_{s_1}^{s_2}ds\frac{\left[\abs{\sin(\theta)}\sinh(2s)-2b\cos(\theta)\cosh(s)\right]}{b(z-y\cos(\theta))-e^{-s}q\abs{\sin(\theta)}},
\end{gather}
	where $s_1$ and $s_2$ are the solutions of Eq.\,(\ref{eq:s_def}) for $x=0$ and $x=Q_c$ respectively. From Eq.\,(\ref{eq:s_integral}) we make the second substitution $t=e^{-s}$ and perform the integration. After a straightforward calculation we find
	\begin{gather}
		\Pi(\vb{q},\omega) = D(E_F)\int_0^{2\pi}d\theta \sum_{b=\pm 1}\sum_{i = 1,5} f_i(\theta,y,z),\\
		f_1(\theta,y,z)= \frac{1}{4}\left(Q_c-y-\sqrt{Q_c^2+2bqQ_c+y^2}\right)\\
		f_2(\theta,y,z)= \frac{b}{4}\left((z+y\cos(\theta))-\frac{y^3\sin^2\theta(2z\cos\theta-y(1+\cos^2\theta))}{(z-y\cos\theta)^3}\right)\ln\left(\frac{y-bz}{\sqrt{Q_c^2+2byQ_c\cos\theta+y^2}-Q_c-bz}\right)\\
		f_3(\theta,y,z)= \frac{by^3\sin^2\theta}{4(z-y\cos\theta)^3}\left[y(1+\cos^2\theta)-2z\right]\ln\left(\frac{y-by\cos\theta}{\sqrt{Q_c^2+2byQ_c+y^2}-Q_c-by\cos\theta}\right)\\
		f_4(\theta,y,z)= \frac{1}{4}\left[\frac{2y\cos\theta}{z-y\cos\theta}-\frac{y^2\sin^2\theta}{(z-y\cos\theta)^2}\right]\left[\frac{y^2\sin^2\theta}{(y-by\cos\theta)}-\frac{y^2\sin^2\theta}{\sqrt{Q_c^2+2byQ_c+y^2}-Q_c-by\cos\theta}\right]\\
		f_5(\theta,y,z)=-\frac{by^2\sin^2\theta}{8(z-y\cos\theta)}\left[\frac{y^2\sin^2\theta}{(y-by\cos\theta)^2}-\frac{y^2\sin^2\theta}{(\sqrt{Q_c^2+2byQ_c+y^2}-Q_c-by\cos\theta)^2}\right]
	\end{gather}

It turns out that the polarization is almost entirely determined by $f_4+f_5$. For the remainder of this note, we work under this assumption.

Let us first focus on the terms in $f_4$ and $f_5$ that do not depend on the upper bound of the radial integral $Q_c$. We refer to these terms as $f_{4A}$ and $f_{5A}$. Summing over the parameter $b$, we can write
\begin{gather}\label{eq:f4A}
	f_{4A} = \frac{y}{2}\lrb{\frac{2y\cos\theta}{z-y\cos\theta}-\frac{y^2\sin^2\theta}{(z-y\cos\theta)^2}},\\
	f_{5A} = -\frac{y^2\cos(\theta)}{2(z-y\cos(\theta))},\label{eq:f5A}
\end{gather} 
which when added together simplifies to
\begin{equation}\label{eq:fA}
	f_{4A}+f_{5A} \equiv f_{A} = \frac{y}{2}\lrb{\frac{y\cos\theta}{z-y\cos\theta}-\frac{y^2\sin^2\theta}{(z-y\cos\theta)^2}}.
\end{equation}
It turns out that when integrating over the angle $\theta$, the two terms in Eq.\,(11) cancel each other exactly. 

Let us now return to those terms in $f_4$ and $f_5$ that depend on the upper bound $Q_c$. We refer to these as $f_{4B}$ and $f_{5B}$. We assume that $x_s\ll y,z\ll1$. This allows us to expand out the remaining terms using
\begin{equation}
	\sqrt{Q_c^2+2byQ_c\cos(\theta)+y^2}-Q_c-by\cos(\theta)\approx \frac{y^2}{2Q_c}\sin^2\theta\left(1-b\frac{y}{q_c}\cos\theta\right),
\end{equation}
which is valid for $q\ll q_c\sim 1$. Summing over $b$ and using this expansion, we can write
\begin{gather}
	f_{4B}=-Q_c\lrb{\frac{2y\cos\theta}{z-y\cos\theta}-\frac{y^2\sin^2\theta}{(z-y\cos\theta)^2}}\\
	f_{5B} = \frac{Q_c^2}{2(z-y\cos\theta)}\lrb{\frac{1}{(1-\frac{y}{Q_c}\cos\theta)^2}-\frac{1}{(1+\frac{y}{Q_c}\cos\theta)^2}}	\approx\frac{2yQ_c\cos\theta}{(z-y\cos\theta)}.
\end{gather}
Combining these expressions, we see that 
\begin{equation}
	\Pi(\vb{q},\omega)=D(E_F)\int_{0}^{2\pi}\frac{d\theta}{2\pi}\frac{Q_cy^2\sin^2\theta)}{(z-y\cos\theta)^2}
\end{equation}

We are interested in the plasmon dispersion $\omega(q)$ obtained by solving the equation
\begin{equation}\label{eq:PlasmonEquation}
	1=U(\vb{q})\Pi(\vb{q},\omega),
\end{equation}
where $U(q)\approx(e^2d/\kappa \varepsilon_0)$ is the Coulomb interaction in the graphene-dielectric-metal stack. We assume that the dispersion is linear, i.e. $\omega=c_s v_F\abs{q}$ and we solve for the slope $c_s$. Let us assume that $q_s\ll1$, and expand the velocity as $c_s=c_{s0}+\delta c_{s1}+\delta c_{s2}$, where $c_{s0}$ is the solution of Eq.\,(\ref{eq:PlasmonEquation}) with $q_s=0$, and $\delta c_{s1(s2)}$ is the first (second) order correction in $q_s$. To obtain this, we expand $Q_c$ in powers of $q_s$. We find  
\begin{gather}
	Q_c(\theta)=1-x_s\cos(\theta+\theta_q)-\frac{x_s^2}{2}\sin^2(\theta-\theta_q),\\
	\Pi(\vb{q},\omega)=D(E_F)\left[\Pi_0(\vb{q},\omega)+\delta\Pi_1(\vb{q},\omega)+\delta\Pi_2(\vb{q},\omega)\right],\\
	\Pi_0(\vb{q},\omega)=\int_{0}^{2\pi}\frac{d\theta}{2\pi}\frac{y^2\sin^2\theta}{(z-y\cos\theta)^2},\\
	\Pi_1(\vb{q},\omega)=x_s\int_{0}^{2\pi}\frac{d\theta}{2\pi}\frac{y^2\sin^2\theta\cos(\theta+\theta_q))}{(z-y\cos(\theta))^2},\\
	\Pi_2(\vb{q},\omega)=-\frac{x_s^2}{2}\int_0^{2\pi}\frac{d\theta}{2\pi}\frac{y^2\sin^2\theta\sin^2(\theta+\theta_q))}{(z-y\cos(\theta))^2}.
\end{gather} 
Performing the integration we find 
\begin{gather}
	\Pi_0(\vb{q},\omega)=\left[\frac{c_s}{c_s^2-1}-1\right]\\
	\Pi_1(\vb{q},\omega)=x_s\cos\theta_q\left[2c_s+\frac{1-2c_s^2}{\sqrt{c_s^2-1}}\right]\\
	\Pi_2(\vb{q},\omega)=\frac{x_s^2}{2}\left[\cos^2\theta_q\left(\frac{3}{2}-3c_s^2+\sqrt{c_s^2-1}\right)-\sin^2\theta_q\left(\frac{1}{2}-3c_s^2+\frac{3c_s^3-c_s}{\sqrt{c_s^2-1}}\right)\right].
\end{gather}
We expand the velocity as $c_s=c_{s0}+\delta c_{s1}+\delta c_{s2}$. Expanding Eq.\,(\ref{eq:PlasmonEquation}) we find
\begin{gather}
	c_{s0}=\sqrt{\frac{(1+A)^2}{1+2A}}\\
	\delta c_{s1} =-q_s\cos\theta_q(c_{s0}^2-1)^{3/2}\lrp{\frac{2c_{s0}^2-1}{\sqrt{c_{s0}^2-1}}-2c_{s0}}\\
	\delta c_{s2} = -\frac{q_s^2}{2}(c_{s0}^2-1)^{3/2}\left[\cos^2\theta_q\left(3c_{s0}^2-\frac{3}{2}-3c_{s0}\sqrt{c_{s0}^2-1}\right)-\sin^2\theta_q\left(\frac{1}{2}-3c_s^2+\frac{3c_s^3-c_s}{\sqrt{c_s^2-1}}\right)\right]
\end{gather}